\documentclass[journal]{IEEEtran}
\usepackage{epsfig,latexsym,amssymb,amsmath,graphics,color}
\usepackage{graphicx,subfigure}
\usepackage{float}
\usepackage{caption}
\usepackage{algorithm}
\usepackage{algpseudocode}

\newtheorem{prop}{Proposition}

\newtheorem{cor}{Corollary}

\newtheorem{lm}{Lemma}

\newtheorem{thm}{Theorem}

\newcommand{\be}{\begin{eqnarray}}
\newcommand{\ee}{\end{eqnarray}}
\newcommand{\benn}{\begin{eqnarray*}}
\newcommand{\eenn}{\end{eqnarray*}}
\def\IR{\rm I \kern-0.20em R}
\newcommand{\utwi}[1]{\mbox{\boldmath $ #1$}}

\newcommand{\bthm}{\begin{thm}}
\newcommand{\ethm}{\end{thm}}

\newcommand{\bcor}{\begin{cor}}
\newcommand{\ecor}{\end{cor}}
\newcommand{\bprop}{\begin{prop}}
\newcommand{\eprop}{\end{prop}}
\newcommand{\blm}{\begin{lm}}
\newcommand{\elm}{\end{lm}}
\newcommand{\beq}{\begin{equation}}
\newcommand{\eeq}{\end{equation}}
\newcommand{\ber}{\begin{eqnarray}}
\newcommand{\eer}{\end{eqnarray}}

\newcommand{\bproof}{\begin{proof}}
\newcommand{\eproof}{\end{proof}}

\newcommand{\bit}{\begin{itemize}}
\newcommand{\eit}{\end{itemize}}
\newcommand{\ben}{\begin{enumerate}}
\newcommand{\een}{\end{enumerate}}
\newcommand{\bdesc}{\begin{description}}
\newcommand{\edesc}{\end{description}}
\newcommand{\beqarrn}{\begin{eqnarray*}}
\newcommand{\eeqarrn}{\end{eqnarray*}}
\newcommand{\bproofof}{\begin{proofof}}
\newcommand{\eproofof}{\end{proofof}}
\newenvironment{rem}{\begin{trivlist}\item[]{\bf
Remark:}\hspace{4mm}}{\end{trivlist}}
\newcommand{\brem}{\begin{rem}}
\newcommand{\erem}{\end{rem}}
\newenvironment{rems}{\begin{trivlist}\item[]{\bf
Remarks}\begin{itemize}}{\end{itemize}\end{trivlist}}
\newcommand{\brems}{\begin{rems}}
\newcommand{\erems}{\end{rems}}
\newtheorem{fact}{Fact}
\newcommand{\bfact}{\begin{fact}}
\newcommand{\efact}{\end{fact}}
\newtheorem{examp}{Example}
\newcommand{\bexamp}{\begin{examp}\rm}
\newcommand{\eexamp}{\end{examp}}
\newtheorem{defn}{Definition}
\newcommand{\bdefn}{\begin{defn}\rm}
\newcommand{\edefn}{\end{defn}}

\newtheorem{alg}{Algorithm}
\newcommand{\balg}{\begin{alg}}
\newcommand{\ealg}{\end{alg}}

\newtheorem{prob}{Problem}
\newcommand{\bprob}{\begin{prob}}
\newcommand{\eprob}{\end{prob}}

\newcommand{\bvtm}{\begin{verbatim}}
\newcommand{\bfig}{\begin{figure}}
\newcommand{\efig}{\end{figure}}
\newcommand{\bcen}{\begin{center}}
\newcommand{\ecen}{\end{center}}

\long\def\comment#1{}

\def \n2{{N_0 \over 2}}

\def \h5{\hspace{0.5in}}

\newcommand{\bc}{{\utwi{c}}}

\newcommand{\bh}{{\utwi{h}}}

\newcommand{\bn}{{\utwi{n}}}

\newcommand{\br}{{\utwi{r}}}
\newcommand{\bs}{{\utwi{s}}}

\newcommand{\bx}{{\utwi{x}}}
\newcommand{\by}{{\utwi{y}}}

\newcommand{\bC}{{\utwi{C}}}
\newcommand{\bD}{{\utwi{D}}}
\newcommand{\bE}{{\utwi{E}}}
\newcommand{\bF}{{\utwi{F}}}

\newcommand{\bH}{{\utwi{H}}}
\newcommand{\bI}{{\utwi{I}}}

\newcommand{\bS}{{\utwi{S}}}

\newcommand{\bU}{{\utwi{U}}}
\newcommand{\bV}{{\utwi{V}}}
\newcommand{\bW}{{\utwi{W}}}

\def\IR{\mathbb R}

\renewcommand{\baselinestretch}{1.0}

\input{epsf}
\begin{document}
\title{SCMA with Low Complexity Symmetric Codebook Design for Visible Light Communication}

\author{Shun Lou, Chen Gong, Qian Gao, and Zhengyuan Xu
\thanks{This work was supported by National Key Basic Research Program of China
(Grant No. 2013CB329201), Key Program of National Natural Science Foundation
of China (Grant No. 61631018), Key Research Program of Frontier
Sciences of CAS (Grant No. QYZDY-SSW-JSC003), Key Project in Science
and Technology of Guangdong Province (Grant No. 2014B010119001), Shenzhen
Peacock Plan (No. 1108170036003286), and the Fundamental Research
Funds for the Central Universities.
The authors are with Key Laboratory of Wireless-Optical Communications,
Chinese Academy of Sciences, University of Science and Technology of
China, Hefei, Anhui 230027, China. Z. Xu is also with Shenzhen Graduate
School, Tsinghua University, Shenzhen 518055, China. Email: loushun@mail.ustc.edu.cn; \{cgong821, qgao,xuzy\}@ustc.edu.cn.}}

\markboth{}%
{Shell \MakeLowercase{\textit{et al.}}: Bare Demo of IEEEtran.cls for Journals}
\maketitle
\begin{abstract}
Sparse code multiple access (SCMA) is attracting significant research interests currently, which is considered as a promising multiple access technique for 5G systems. It serves as a good candidate for the future communication network with massive nodes due to its capability of handling user overloading. Introducing SCMA to visible light communication (VLC) can provide another opportunity on design of transmission protocols for the communication network with massive nodes due to the limited communication range of VLC, which reduces the interference intensity. However, when applying SCMA in VLC systems, we need to modify the SCMA codebook to accommodate the real and positive signal requirement for VLC. We apply multi-dimensional constellation design methods to SCMA codebook. To reduce the design complexity, we also propose a symmetric codebook design. For all the proposed design approaches, the minimum Euclidean distance aims to be maximized. Our symmetric codebook design can reduce design and detection complexity simultaneously. Simulation results show that our design implies fast convergence with respect to the number of iterations, and outperforms the design that simply modifies the existing approaches to VLC signal requirements.
\end{abstract}

\IEEEpeerreviewmaketitle
\section{Introduction}
In recent years, the 5th generation communication (5G) is attracting more and more attention due to the rapid development of mobile communication \cite{osseiran2014scenarios}. The main purpose for 5G communication is to support three typical scenarios including enhanced mobile broadband (eMBB), massive machine type communications (mMTC) and ultra-reliable and low latency communications (uRLLC). The conventional orthogonal multiple access (OMA) schemes such as orthogonal frequency division multiplexing (OFDM) exhibit certain bottleneck in coping with massive machine connectivity. Then, non-orthogonal multiple access such as low-density signature (LDS) is proposed \cite{hoshyar2008novel} \cite{van2009multiple}, which can support massive connectivity for massive number of users.

Sparse code multiple access (SCMA) with optimized codebook design can outperform LDS due to the shaping gain. SCMA is first proposed as an improvement of LDS in \cite{nikopour2013sparse}, both having low density characteristic. It is shown in \cite{zhang2017poc} that SCMA can support massive connectivity, low latency and reduced overhead for grant-free uplink. SCMA provides the opportunity of transmitting the amount of data more than that of resource block called overload. An SCMA encoder maps the information bits to a multi-dimensional codeword, which is different from the conventional modulation. Note that the transmitted codeword is sparse for mitigating the interference of different users. At the receiver, message pass algorithm (MPA) can be applied to detect the signal of each user \cite{yang2017low} \cite{mu2015fixed}, which can utilize the sparse property of the transmitter, and thus sharply reduce the complexity of multi-user detection. Therefore, it is crucial to design good codebook of each user for a large shaping gain and low complexity detection algorithm.

A multi-stage design approach of SCMA codebooks is proposed in \cite{taherzadeh2014scma}. The mother constellation with a good Euclidean distance profile is created primarily based on the design principles of lattice constellations \cite{boutros1996good}, and then different constellation operators such as phase rotations can be applied on the mother constellation to build multiple codebooks for different users. Finally, all codebooks are mapped into a higher dimension through Latin matrix for sparsity. Multi-dimensional SCMA codebook design based on constellation rotation and interleaving is investigated in \cite{cai2016multi}, as an extension of \cite{taherzadeh2014scma}. In \cite{yu2015optimized}, an improved method based on star-QAM signaling constellations is proposed for designing SCMA codebooks. The aforementioned methods are based on a multi-stage heuristic optimization approach, which may lead to a suboptimal solution. To further improve the performance, a joint optimization of the constellation with mapping matrix is proposed in \cite{peng2017joint}. It solves the optimization problem through semi-definite relaxation (SDR), which outperforms the multi-stage optimization.

On the other hand, visible light communication (VLC) based on light-emitting diode (LED) is attracting more and more attention \cite{jovicic2013visible} \cite{gao2015dc} due to its long lifetime, low power consumption, and the capability for high-rate on-off switching that enables the modulation for communication. Besides the high data rate communication, it is a good candidate for mMTC or uRLLC, with massive communication nodes and low data rate, due to the limited communication range and thus lower interference. Then designing SCMA for a VLC system \cite{feng2016applying} is of significant interest. However, since the signals for LED-based VLC must be real and positive, certain modifications on the conventional SCMA are needed. Note that SCMA is a special type of non-orthogonal multiple access (NOMA), where power-domain NOMA has been studied in \cite{marshoud2016non}. We address the codebook design for SCMA and propose a symmetric codebook structure to reduce the design complexity. Moreover, we formulate the codebook design problem to maximize the minimum Euclidean distance in the codebook, and transform it into a convex second-order cone programming (SOCP) problem. Our symmetric codebook design can significantly reduce the design and detection complexity simultaneously. Simulation results show that our scheme can outperform the design that simply modifies the conventional scheme for non-negative signals.

\section{System Model}
\subsection{Sparse Code Multiple Access}
\begin{figure}[H]
\centering
\includegraphics[width=3.4in]{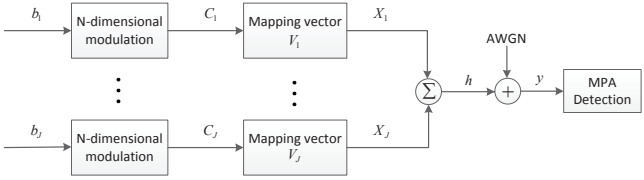}
\caption{SCMA system model.}
\label{fig.scma_system}
\end{figure}
SCMA is a novel codebook-based non-orthogonal multiple access technique that can improve the spectral efficiency. We consider an SCMA system with $J$ users and $K$ physical resources, where the overload factor is defined as $\lambda=J/K$. An SCMA transmitter consists of an SCMA encoder and an SCMA multiplexer, as shown in Figure \ref{fig.scma_system}. The SCMA encoder is defined as a mapping from input bits to a multi-dimensional codeword. In other words, $log_2(M)$ bits are encoded to a $K$-dimensional codeword from the predefined user codebook with size $M$. The $K$-dimensional codeword can be obtained from an $N$-dimension codeword through inserting $K-N$ zeros, which can be regarded as being sparse since $N<K$. The transform process can be regarded as a mapping matrix $\bV$ with $K-N$ zeros in each row. The structure of SCMA can be represented as a factor graph, as shown by Figure \ref{fig.factor_graph}, where RNs and VNs denote resource nodes and variable nodes, respectively.
\begin{figure}[H]
\centering
\includegraphics[width=3.2in]{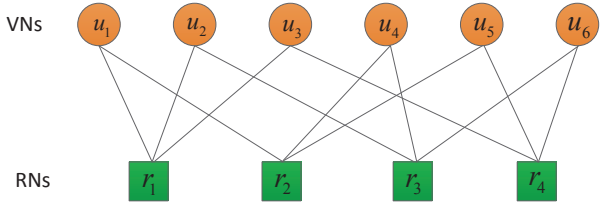}
\caption{A factor graph for SCMA with $J$ = 6 and $K$ = 4.}
\label{fig.factor_graph}
\end{figure}
The corresponding factor graph matrix $\bF$ for SCMA can be expressed as follows,
\begin{equation}\label{graph_matrix}
 \bF =
  \left[
  \begin{array}{cccccc}
          1 & 1 & 1 & 0 & 0 & 0 \\
          1 & 0 & 0 & 1 & 1 & 0 \\
          0 & 1 & 0 & 1 & 0 & 1 \\
          0 & 0 & 1 & 0 & 1 & 1
  \end{array}
  \right],
\end{equation}
where user $j$ and resource $k$ are connected if and only if $F_{j,k}=1$. Let $\bF_i$ denote $i$-th column of $\bF$. Then, the mapping matrix for user $i$ can be obtained via setting $\bV_i$$=diag(\bF_i)$ and eliminating all 0 column vectors, such as
\begin{equation}\label{mapping_matrix}
 \bV_1 =
  \left[
  \begin{array}{cc}
          1 & 0  \\
          0 & 1 \\
          0 & 0  \\
          0 & 0
  \end{array}
  \right]  , \
   \bV_2 =
  \left[
  \begin{array}{cc}
          1 & 0  \\
          0 & 0 \\
          0 & 1  \\
          0 & 0
  \end{array}
  \right] , \
     \bV_3 =
  \left[
  \begin{array}{cc}
          1 & 0  \\
          0 & 0 \\
          0 & 0  \\
          0 & 1
  \end{array}
  \right]
\end{equation}

Since six columns can support six users in the four slots, the above matrix $\bF$ can achieve the overloading factor $\lambda=150\%$. Note that the number of users superposed on a resource expressed by $d_f$ remains the same, and all users have the same number of dimension of non-zero entries $N$. In this case, we have $d_f = 3$ and $N=2$. Then SCMA codewords of $J$ users can be multiplexed in a synchronous manner, defined as SCMA multiplexer, as shown Figure. \ref{fig.sum_user}.
\begin{figure}[H]
\centering
\includegraphics[width=3.0in]{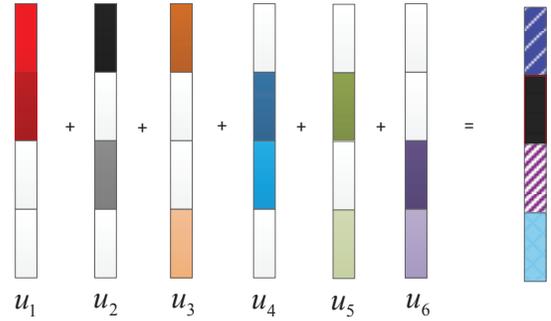}
\caption{SCMA multiplexer with $J$ = 6 and $K$ = 4.}
\label{fig.sum_user}
\end{figure}

The received signal for the SCMA system can be expressed as follows,
\begin{equation}\label{SCMA_multiplexing}
\begin{aligned}
 \by &= \sum_{j=1}^Jdiag(\bh_j)\bx_j +\bn  \\
 &= \sum_{j=1}^Jdiag(\bh_j){\bV_j}\bc_j +\bn,
\end{aligned}
\end{equation}
where $\bh_j$ denotes the channel vector of user $j$; $\bn$ denotes the additive white Gaussian noise (AWGN); $\bc_j$ denotes the $N$-dimensional codeword; and $\bx_j$ denotes the corresponding $K$-dimensional codeword. At the receiver, based on the received signal $\by$ and channel state information (CSI), the detection of $J$ users can be realized by applying MPA, which is an iterative detection approach based on factor graph. The sparsity of codebook may limit the number of connectivity, implying the reduced complexity decoding via message passing even if for a large overloaded factor $\lambda$.

\subsection{Sparse Code Multiple Access for VLC}
Since VLC systems are based on intensity modulation and direct detection (IM-DD), the modulated signals need to be real and positive. Fundamental problems for SCMA are the codebook design and low complexity detection algorithm. Note that the SCMA encoder directly maps the input bits of each user to a multi-dimensional complex codeword encoded according to the predefined codebooks.

However, SCMA for RF communication cannot be applied to VLC systems directly, since the complex signal cannot be transmitted and received in a VLC system, via constructing the sparsity in the frequent domain. An alternative method is to divide the complex signal to real and imaginary parts, and the two parts are transmitted in different resource blocks. The positive signal can be achieved by adding a DC bias. Then we can combine the real and imaginary parts at the receiver to detect the information bits. We can also adopt ACO-OFDM \cite{armstrong2006power} and DCO-OFDM \cite{gonzalez2005ofdm} to transmit real signal by Hermitian symmetry in a VLC system. The SCMA codebook design based on OFDM can be modified to satisfy the non-negative constraint for the VLC. However, the aforementioned methods are both based on structured codebook design for RF technology. Further performance improvement can be achieved if we can design the multi-dimension real codebook for VLC, while maintaining the sparsity.

\section{constellation design}
\subsection{Multi-dimension Constellation Design for SCMA}
For two matrices with the same number of rows $\bU\in \mathbb{C}^{p\times n}$ and $\bW\in \mathbb{C}^{p\times l}$, the set of the arbitrary column sum of $\bU$ and $\bW$ can be defined as follows,
\begin{equation}\label{sum_UW}
\bU\oplus \bW = \bU\otimes \boldsymbol{e}_l^T  + \boldsymbol{e}_n^T\otimes \bW,
\end{equation}
where $\boldsymbol{e}_l^T$ and $\boldsymbol{e}_n^T$ are all-one row vectors with length $l$ and $n$, respectively, and operator $\otimes$ denotes the Kronecker product. The set of the arbitrary column difference of $\bU$ and $\bW$ can be defined as follows,
\begin{equation}\label{minus_UW}
\bU\ominus \bW = \bU\otimes \boldsymbol{e}_l^T  - \boldsymbol{e}_n^T\otimes \bW.
\end{equation}

Let $\bC_i$ denote the predefined codebook for user $i$. The superimposed codeword matrix $\bS$ for all users can be represented as follows,
\begin{equation}\label{sum_C}
\bS = \bC_1\oplus \bC_2\oplus \bC_3\oplus \cdots \oplus \bC_J.
\end{equation}
To improve the detection performance, we aim to maximize the minimum Euclidean distance (MED) in the designed codebook \cite{forney1989multidimensional}. The problem can be formulated as follows,
\begin{equation}\label{optm}
\begin{aligned}
\min_{\bS} \quad &\sum_m\|\bs_m\|^2 \\
s.t. \quad & \|\bs_i-\bs_j\|^2 \geq 1,\quad  1 \leq i\leq j\leq M^J.
\end{aligned}
\end{equation}

The original optimization problem (\ref{optm}) is the nonconvex problem, which is generally difficult to solve. However, there are some good design methods for such optimization problem given in \cite{beko2012designing}. Such optimization problem can be relaxed to a convex second-order cone programming (SOCP) problem. We reshape matrix $\bS$ as a vector $\br$ by column. Then the problem can be reformulated as
\begin{equation}\label{optm_1}
\begin{aligned}
\min \quad &\|\br\|^2 \\
s.t. \quad & 2\br_0^T\bE_{ij}\br - \br_0^T\bE_{ij}\br_0 \geq 1,\quad   1\leq i\leq j\leq M^J,
\end{aligned}
\end{equation}
where $\br_0$ denotes the initial feasible solution; $\bE_{ij}=\bE_i^T\bE_i-\bE_i^T\bE_j-\bE_j^T\bE_i+\bE_j^T\bE_j$, $\bE_i = \boldsymbol{u}_i^T\otimes \bI_N$; $\boldsymbol{u}_i$ represents the $i$-th column of identity matrix $\bI_M$; and $\bI_N$ denotes an $N$ order identity matrix. Note that the elements in $\br$ need to be masked by the mapping matrix via setting the corresponding positions to be zero, in order to guarantee sparsity. We can guarantee $\br^T\bE_{ij}\br\geq 1$ since $(\br-\br_0)^T\bE_{ij}(\br-\br_0)\geq 0$. The result $\|\br\|^2\leq\|\br_0\|^2$ is obvious since $\br$ is a feasible solution to optimization problem (\ref{optm_1}). The codebook design can be conducted via iteratively setting $\br$ to be $\br_0$ and solving optimization problem (\ref{optm_1}). The procedure for solving the multi-dimension constellation design for SCMA is shown in Algorithm $1$.
\begin{algorithm}[h] \label{SOCP_algorithm}
\caption{Iterative solution to codebook design}
\begin{algorithmic}[1]
\State Random generate sparse $\br_0$ from the mapping matrix;
\If {$\br_0^T\bE_{ij}\br_0\leq 1$;}
\State go to step $1$;
\EndIf
\State Solve optimization problem (\ref{optm_1}), and obtain solution $\br^*$;
\If {$\|\br^*-\br_0\|\geq 0.01$;}
\State Set $\br_0=\br^*$, and go to step $5$;
\EndIf
\State Return $\br^*$ as the output codebook design.
\end{algorithmic}
\end{algorithm}

The transmitted signals of each user are multiplexed based on different predefined codebooks, and the design complexity grows exponentially with the number of users. Therefore, the main challenge of codebook design for SCMA is to decrease the complexity due to the multiplexing of different users' codebooks.

\subsection{Reduced Complexity Design for Symmetric Constellation}
We define $\bD_i = \bC_i\ominus \bC_i$ as difference of arbitrary two codewords in the codebook for user $i$, and merge the repeated column vectors of matrix $\bD_i$. Therefore matrix $\bD$ for the difference between arbitrary two column vectors $\bs_i-\bs_j$ from $\bS$ can be represented as the following direct sum of matrices,
\begin{equation}\label{sum_D}
\begin{aligned}
\bD = \bD_1\oplus \bD_2\oplus \bD_3\oplus \cdots \oplus \bD_J.
\end{aligned}
\end{equation}

We aim to obtain reduced complexity design via introducing certain symmetry to the codebook. Consider a simple example of four codewords in the codebook for each user. Let $c_{ij}$ denote absolute value of signals for user $i$ and resource block $j$. Our design structure is to fix the absolute value of signals, given by $\bc_i = [c_{i1},c_{i2},\cdots,c_{iK}]^T$. Then codebook $\bC_i$ can be represented as
\begin{equation}\label{C_matrix}
\begin{aligned}
     \bC_i =
  \left[
  \begin{array}{cccc}
          1 & -1  &  1   & -1   \\
          1 & -1  & 1    & -1  \\
          1 & -1   & -1   & 1   \\
          1 & -1   &-1    & 1
  \end{array}
  \right]\odot (\bc_i\otimes \boldsymbol{e}_4^T),
\end{aligned}
\end{equation}
where $\odot$ denote the element-wise product. The computational complexity of codebook design can be significantly reduced since the number of variables and constraints to be optimized can be reduced. The difference matrix $\bD_i$ for our symmetric constellation design can be given by
\begin{equation}\label{D_matrix}
\begin{aligned}
     \bD_i =
  \left[
  \begin{array}{ccccccccc}
          0 & 2   &-2   &-2 & 2  &  2   & -2 & 0  &  0   \\
          0 & 2   &-2   &-2 & 2  &  2   & -2 & 0  &  0  \\
          0 & 2   &-2   & 2 & -2  &  0   & 0 & 2  &  -2     \\
          0 & 2   &-2   & 2 & -2  &  0   & 0 & 2  &  -2
  \end{array}
  \right].
\end{aligned}
\end{equation}

A more general case for the codebook design can be conducted as follows. Let $\bD_{ri} = \bD_1\oplus \cdots\oplus \bD_{i-1}\oplus \bD_{i+1}\cdots \oplus \bD_{J}$ denote the direct sum except for $\bD_i$. Then we can get $\bD = \bD_{ri}\oplus \bD_i$. Note that maximizing MED of matrix $\bD_i$ is the subproblem of maximizing MED of matrix $\bD_i$ because there is an all-zero column vector in matrix $\bD_{ri}$. Based on the aforementioned criterion, we propose a Hardmard-based structure for generating the symmetric matrix $\bH_i$ for each user, which can guarantee the MED of matrix $\bC_i$, given by
\begin{equation}\label{H_matrix}
\begin{aligned}
     \bH_{i+1} =
  \left[
  \begin{array}{cccc}
          \bH_i & \bH_i  \\
          \bH_i & -\bH_i
  \end{array}
  \right].
\end{aligned}
\end{equation}
The basic matrix $H_0$ is represented as
\begin{equation}\label{H0_matrix}
\begin{aligned}
     \bH_0 =
  \left[
  \begin{array}{cccc}
          1 & -1  \\
          1 & -1
  \end{array}
  \right].
\end{aligned}
\end{equation}
Then the construction can be achieved via the element-wise product of matrix $\bH_{i+1}$ and $\bc_i\otimes\boldsymbol{e}_h^T$, where $h$ denotes the column number of $\bc_i$.

\subsection{Complexity Analysis for Design and Detection}
Note that the computational complexity mainly results from solving the optimization problem (\ref{optm_1}). Obviously, the design may suffer significant computational complexity if a general structure on the codebook needs to be designed. Consider scheme $1$, where we express the difference as $\bs_i-\bs_j$ for each pair of elements in $\bS$. Then the size of $\bS$ set is $M^J$, and then the number of constraints is given by $\frac{M^J(M^J-1)}{2}$, since each $\bs_i$ needs to be compared with all other columns in the matrix $\bS$. Consider scheme~$2$, where we express the difference as $\bD$ for the conventional structure. Then the size of set $\bD_i$ is given by $P=M(M-1)+1$, since each column of $\bC_i$ needs to be compared with all other columns, which leads to the size of $M(M-1)$, and each column also compares with itself, resulting in identical all-zero vector. Thus we obtain another size of $1$. The complexity can be given by $P^J-1$ since all combinations of $J$ users need to be incorporated, where we need to subtract $1$ from the final size due to removing the all-zero codeword. However, such number is still high for the designing problem. For our proposed scheme, we express the difference as $\bD$ for our symmetric structure. Then the size can be significantly reduced due to sharp decrease of $P$. The computational complexity of the above three schemes is shown as TABLE \ref{complexity_compare} for $M=4$ and $J=6$. We can see that our design scheme can greatly reduce the computational complexity.
\begin{table}[H]
\caption{Complexity measurement.\label{complexity_compare}}
\centering
\begin{tabular}{|c|c|c|c|}
\hline
Scheme                &Number of variables          &$P$                 &Number of constraints          \\ \hline
Scheme $1$              &96                           &$\backslash$         &8386560            \\  \hline
Scheme $2$             &96                           &13                  &4826808            \\  \hline
Proposed scheme            &24                          &9                   &531440               \\
\hline
\end{tabular}
\end{table}

MPA is applied for multiuser detection for SCMA, which has significantly lower complexity than the optimal maximum a posteriori (MAP) decoder, due to the sparsity of codebook that can form a Tanner graph. It is an iterative propagation of message among VNs and RNs, shown as follows,
\begin{equation}\label{MPA_ru}
\begin{aligned}
&I_{r_k\to u_j}^{(t)}(x_j)= \\
&\sum_{\backsim x_j}\Bigg\{{\frac{1}{2\pi\sigma^2}\exp(-\frac{1}{2\sigma^2}\Big\|y_k-\sum_{m\in V(k)}h_{k,m}x_{k,m}\Big\|^2)}\\
&{\prod_{p\in V(k)\backslash j}I_{u_p-r_k}^{t-1}(x_p)}\Bigg\};
\end{aligned}
\end{equation}

\begin{equation}\label{MPA_ur}
\begin{aligned}
I_{u_j\to r_k}^{(t)}(x_j)= \prod_{m\in R(j)\backslash k}I_{r_m\to u_j}^{(t)}(x_j),
\end{aligned}
\end{equation}
where $V(k)$ represents the set of all VNs connected to $k$-th RN $r_k$; $R(j)$ represents the set of all RNs connected to $j$-th VN ${u_j}$; $t$ denotes the number of iteration; and $\backsim{x_j}$ denotes the taking marginal function for variables except $x_j$ for factor graph. Fixing the maximum iteration number $t_{max}$, the decoding output can be expressed as
\begin{equation}\label{output}
\begin{aligned}
Output(x_j) = \prod_{k\in R(j)}I_{r_k\to u_j}^{(t_{max})}(x_j).
\end{aligned}
\end{equation}

Note that sparsity can reduce the decoding complexity for SCMA, while our design structure can further reduce the detection complexity. Note that the decoding complexity of MPA mainly comes from Equation (\ref{MPA_ru}), where for computing the probability of all users' signal combination, we need to compute $\sum_{m\in V(k)}h_{k,m}x_{k,m}$. However, since every resource block only has two different values for each user in our design structure, the complexity of probability calculation can be reduced to $2^J$, compared with $M^J$ for the general approach.

\section{Numerical Results}
Simulation results are presented to compare our design method with conventional schemes via BER performance. The typical parameters for SCMA are shown in TABLE \ref{table4}, same as the parameters adopted in references \cite{taherzadeh2014scma} \cite{cai2016multi} \cite{yu2015optimized}. Note that the number of resources is defined in complex codebook, while we need to double it when replacing a complex codebook with a real codebook, and the factor graph matrix is represented as $[\bF;\bF]$ via matrix repeating. The codewords in the designed codebook is shown in the Appendix.
\begin{table}[H]
\caption{System parameters.\label{table4}}
\centering
\begin{tabular}{|c|c|}
\hline
Parameters   & Value      \\ \hline
Number of resources $K$        &4          \\
Number of all users $J$        &6         \\
Codebook size $M$ of each user    &4         \\
Number of active users $d_r$   &3         \\
Number of active resource $N$   &2           \\
Overload factor $\lambda$      &150\%            \\
Number of Iteration for MPA           &5            \\
Channel model                  &AWGN         \\
Number of frame                &50        \\
Length bit each frame          &1024       \\
\hline
\end{tabular}
\end{table}

The BERs versus SNR for different codebook design approaches are shown in Figure \ref{fig.sim_result}, where LDS \cite{hoshyar2008novel}, shuffling-SCMA \cite{taherzadeh2014scma}, MD-SCMA \cite{cai2016multi}, and starQAM-SCMA \cite{yu2015optimized} are modified to accommodate the real and positive signals for visible light communication. Symmetric-SCMA denotes the symmetric codebook design approach proposed in this paper. Performance improvement of our design is observed compared with the above four benchmark schemes.

\begin{figure}[H]
\centering
\includegraphics[width=3.4in]{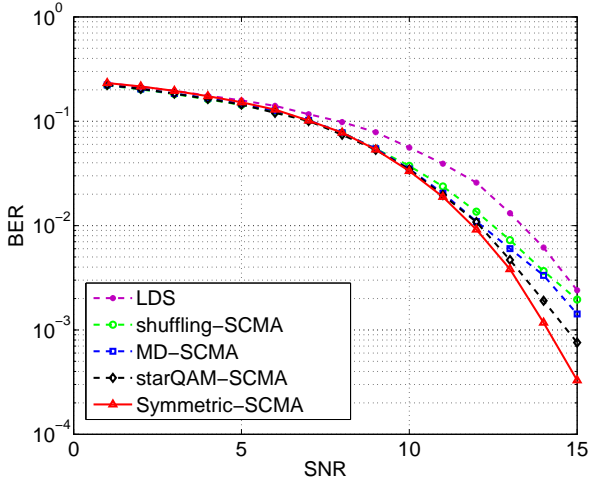}
\caption{BER comparison between the proposed symmetric design and the benchmark design approaches.}
\label{fig.sim_result}
\end{figure}

\begin{figure}[H]
\centering
\includegraphics[width=3.4in]{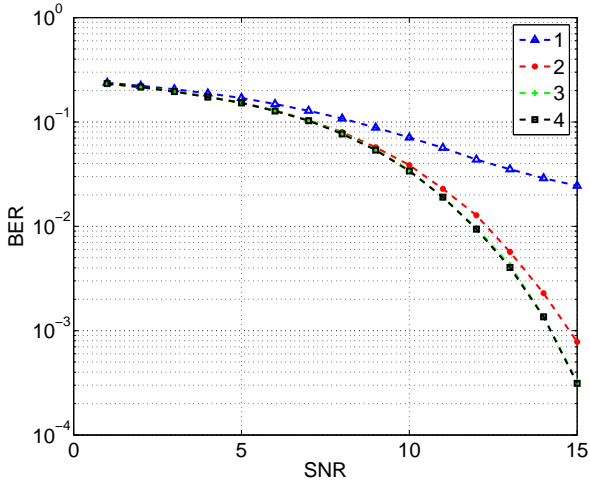}
\caption{The detection bit error probability for different numbers of iterations for MPA decoding.}
\label{fig.iteration}
\end{figure}

Figure \ref{fig.iteration} shows the detection bit error probability for the MPA under different number of iteration for the proposed design structure. Increasing the number of iterations can reduce the detection error probability, and the performance improvement beyond three iterations is smaller than that below three iterations, i.e., three iterations suffice to provide a reasonably good performance. Further increasing the number of iterations can hardly provide additional BER reduction.

The codebook power $\|\br\|^2$ versus the number of iterations for Algorithm $1$ is shown in Figure \ref{fig.eng_iter}. It is seen that after ten iterations, the codebook power becomes saturated and close to zero, which implies that solving the problem (\ref{optm_1}) via ten iterations suffices to provide a good codebook design.

\begin{figure}[H]
\centering
\includegraphics[width=3.4in]{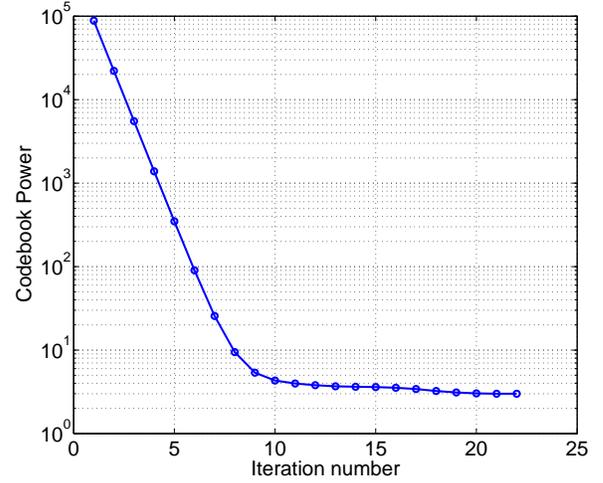}
\caption{The codebook power $\|\br\|^2$ versus the number of iterations for Algorithm $1$.}
\label{fig.eng_iter}
\end{figure}

\section{Conclusions}
In this paper, we have investigated the reduced complexity codebook design for SCMA for VLC systems. We have proposed a symmetric codebook design structure, and optimized the multi-dimensional constellation design, which can maximize the minimum Euclidean distance between codewords in the codebook. We have also proposed symmetric codebook to reduce the design complexity when considering the combination of different users. The proposed symmetric codebook design can reduce the design and detection complexity simultaneously. Simulation results show that our design outperforms the design that simply modifies the existing approaches for visible light communication.

\section{Appendix}
The codebooks used in the simulations for six users are shown as follows,
\begin{equation}
\bC_1=
\left[
\begin{array}{cccc}
    0.7071    &0.7071   &-0.7071   &-0.7071\\
   0.0000   &0.0000    &0.0000    &0.0000\\
    0         &0         &0   &0\\
    0         &0         &0   &0\\
    0.3536   &-0.3536    &0.3536   &-0.3536\\
    0.3536   &-0.3536    &0.3536   &-0.3536\\
    0         &0         &0   &0\\
    0         &0         &0   &0\\
\end{array}
\right].
\end{equation}

\begin{equation}
\bC_2=
\left[
\begin{array}{cccc}
    0.3536    &0.3536   &-0.3536   &-0.3536 \\
    0         &0         &0   &0\\
    0.3536    &0.3536   &-0.3536   &-0.3536\\
    0         &0         &0   &0\\
    0.7071   &-0.7071    &0.7071   &-0.7071\\
    0         &0         &0   &0\\
    0.0000   &-0.0000    &0.0000   &-0.0000\\
    0         &0         &0   &0\\
\end{array}
\right].
\end{equation}

\begin{equation}
\bC_3=
\left[
\begin{array}{cccc}
   0.0000   &0.0000    &0.0000    &0.0000\\
    0         &0         &0   &0\\
    0         &0         &0   &0\\
    0.5000    &0.5000   &-0.5000   &-0.5000\\
    0.3536   &-0.3536    &0.3536   &-0.3536\\
    0         &0         &0   &0\\
    0         &0         &0   &0\\
    0.3536   &-0.3536    &0.3536   &-0.3536
\end{array}
\right].
\end{equation}

\begin{equation}
\bC_4=
\left[
\begin{array}{cccc}
    0         &0         &0   &0\\
    0.3536    &0.3536   &-0.3536   &-0.3536\\
    0.3536    &0.3536   &-0.3536   &-0.3536\\
    0         &0         &0   &0\\
    0         &0         &0   &0\\
    0.0000   &0.0000    &0.0000   &0.0000\\
    0.5000   &-0.5000    &0.5000   &-0.5000\\
    0         &0         &0   &0\\
\end{array}
\right].
\end{equation}

\begin{equation}
\bC_5=
\left[
\begin{array}{cccc}
    0         &0         &0   &0\\
    0.6036    &0.6036   &-0.6036   &-0.6036  \\
    0         &0         &0   &0\\
    0.2500    &0.2500   &-0.2500   &-0.2500\\
    0         &0         &0   &0\\
    0.7071   &-0.7071    &0.7071   &-0.7071\\
    0         &0         &0   &0\\
    0.0000   &0.0000   &0.0000    &0.0000
\end{array}
\right].
\end{equation}

\begin{equation}
\bC_6=
\left[
\begin{array}{cccc}
    0         &0         &0   &0\\
    0         &0         &0   &0\\
    0.7071    &0.7071   &-0.7071   &-0.7071\\
    0.0000    &0.0000    &0.0000    &0.0000\\
    0         &0         &0   &0\\
    0         &0         &0   &0\\
    0.2500   &-0.2500    &0.2500   &-0.2500\\
    0.6036   &-0.6036    &0.6036   &-0.6036\\
\end{array}
\right].
\end{equation}

\bibliography{reference}
\end{document}